\pdfoutput=1

\documentclass[aps,preprintnumbers,amsmath,amssymb,twocolumn, tightenlines,superscriptaddress]{revtex4}
\usepackage{graphicx}
\usepackage{slashed}
\usepackage{tikz,mathpazo}
\usetikzlibrary{shapes.geometric, arrows}

\begin{document}


\title{Local suppression and enhancement of pairing condensate under rotation}

\author {Lingxiao Wang}
\address{Physics Department, Tsinghua University, , Beijing 100084, China.}
\author {Yin Jiang} 
\address{Physics Department, Beihang University, 37 Xueyuan Rd, Beijing 100191, China}
\author {Lianyi He}
\address{Physics Department, Tsinghua University, , Beijing 100084, China.}
\author {Pengfei Zhuang}
\address{Physics Department, Tsinghua University, , Beijing 100084, China.}
\date{\today}

\begin{abstract}
The rotation induced inhomogeneous problem is non-trivial and inevitable. In this paper a generic framework is developed to investigate the inhomogeneous condensate in a system of fermions under the presence of rotation. It is a self-consistent method basing on a set of relativistic BdG equations solved with typical iteration algorithm. Taking the chiral condensate for example we study rotational effects numerically and discover two inhomogeneous effects, the local rotational suppression effect and centrifugal effect. They may have significant impacts on the phase structure of various kinds of matter. Several systems in different physics branches have been discussed in the paper.
\end{abstract}

\maketitle

{\it Introduction.---}
Inhomogeneous phenomena emerging in rotating matter have attracted a lot of interest in different branches of physics. In cold atom physics, the rotation would not only induce significant modifications on phase structures\cite{Fetter:2009zz} but also excite novel inhomogeneous states\cite{Matthews:1999zz,Madison:2000zz,Haljan:2001zz}, such as the topological nontrivial vortex state\cite{Eto:2013hoa}. In astrophysics, neutron star has kept people a long-last interest on its spinning effect for decades\cite{Watts:2016uzu,Ozel:2016oaf}. Besides the usual structure studies which related to the phase identification in each layer, pulsar, which rotating rapidly and persistently, shows more mysterious properties, such as the glitch behavior which may indicate the rotational speed distribution among different layers and coupling of them\cite{Anderson:1975zze,Link:1992mdl,Melatos:2007px,Chamel:2012ae}. In the high-energy nuclear physics, motivated by the new record of vorticity on LHC, i.e. around 0.006GeV\cite{STAR:2017ckg}, more studies have been performed to explore both the vorticity distribution\cite{Liang:2004ph,Becattini:2007sr,Csernai:2013bqa,Jiang:2016woz,Xia:2018tes,Wei:2018zfb} and its impacts on the phase structure of the dense and hot strongly interacting Quantum Chromodynamics(QCD) matter\cite{Yamamoto:2013zwa,Chen:2015hfc,Jiang:2016wvv,Ebihara:2016fwa,Chernodub:2016kxh,Chen:2017xrj,Liu:2017spl,Wang:2018sur,Huang:2017pqe}. According to simulations of transport models\cite{Jiang:2016woz,Xia:2018tes,Wei:2018zfb} the lifetime of the vorticity is relatively long and the spatial fluctuation is very large, i.e. in the range of 0$\sim$0.2GeV in the early stage of the evolution of quark gluon plasma(QGP) fireball produced in relativistic heavy ion collision(HIC). Hence it is tempting to ask what kind of influence would the inhomogenousness of vorticity bring on the phase structure and the parton transport in QGP. Recently a series of novel chiral transport phenomena, such as chiral magnetic/vortical effects(CME/CVE)\cite{Kharzeev:2007tn,Son:2009tf,Kharzeev:2010gr,Kharzeev:2007jp} and chiral magnetic/vortical wave(CMW/CVW)\cite{Kharzeev:2010gd,Burnier:2011bf,Jiang:2015cva}, are of great interest as potential determinate signals of chiral restoration in QGP. Therefore studying the inhomogeneous rotational effects self-consistently is necessary for characterizing these chiral transport phenomena quantitatively.

Apart from phenomenological applications in above, there are also theoretical interests on the rotation induced inhomogeneous condensate which usually serves as order parameter of continues phase transition. The upper limit of speed forbids an infinite rotating system with a uniform and finite angular velocity. Hence at least one of two challenges one shall confront, the finite-size problem or the nonuniform rotation speed(decaying towards zero at infinity). Under local density approximation(LDA), which assumes the system varies slowly with position, the first one is easy to solve in the uniform rotation case. Because by neglecting derivatives with respect to the position the problem will be reduced to solving a usual gap equation.  Works on this simplified system have been done in \cite{Jiang:2016wvv,Ebihara:2016fwa,Chen:2017xrj}. However it is shown that fermion pairings shall still depend on the radial coordinate explicitly. Furthermore the choice of boundary conditions has even been demonstrated to be decisive on the existence of condensates. Although invalid for a generic rotating system, these tentative studies suggest that the inhomogeneous problem is inevitable in rotation relevant topics. Conceptually it is also natural because the centrifugal effect should always exist in a nonrigid system.

The pairing phenomenon between fermions under suitable conditions, which serves as an important mechanism of phase transitions, encompasses a wide range of systems. In this report we will focus on a generic relativistic system of spinor fermion under medium rotation. In order to solve it self-consistently the Bogoliubov–de Gennes(BdG) method\cite{Nickel:2008ng,Iskin:2012rso,Buballa:2014tba} is explanted into a relativistic version. The original BdG method is a reliable numerical algorithm for studying inhomogeneous problems in the non-relativistic physics, such as cold atom systems\cite{Tewari:2007,Bausmerth:2008,Sato:2009}. It is a typical iteration algorithm of solving the Sch\"{o}rdinger equation whose Hamiltonian is a functional of the set of target eigenstates. Starting from a speculated Hamiltonian and expanding the target eigenstates with a proper basis, the Sch\"{o}rdinger eigen equation could be converted to a set of algebra equations of those expanding coefficients. And it is solved self-consistently by iteration, that is modifying the solution in the next step with the previous one until they are consistent with each other. In this work the BdG method is reformulated to solve the Dirac equation in 2+1D system. The dimension choice could not only simplify the numerical computation but also eliminate the subtlety arisen from regularization scheme choices of the four-fermion interaction model. Because such an interaction is renormalizable in 2+1D case\cite{Appelquist:1986fd,Rosenstein:1990nm}. And as an analogy to the 3+1D chiral condensate, the scalar pairing $\langle\bar\psi  \psi\rangle$ is studied as a example. It is worthwhile to emphasize that in principle any type of two fermions pairing could be straightforwardly studied in this framework, such as diquark, pion condensate and pairing states with nonzero angular momentum. In the following $\langle\bar\psi  \psi\rangle$ will be also referred as chiral condensate in analogy to the usual 3+1D system, although the chirality definition is a little tricky in the 2+1D system.

{\it A Self-Consistent Framework.---}
We adopt the Nambu–-Jona-Lasinio(NJL) model to study a relativistic system of fermion under nonuniform rotation in 2+1D. The angular velocity is perpendicular to the 2D plane along z-direction. As our previous work we introduce the rotation via the curved metrics and expand the Lagrangian density to the linear order of velocity $\mathcal{O}(v)$. In mean field approximation it is reduced to
\begin{eqnarray}
\mathcal{L}&&=\bar{\psi}i\slashed{\tilde D}{\psi}+\frac{G}{2}\left[(\bar{\psi}{\psi})^2+(\bar{\psi}i{\tilde\gamma}_5{\psi})^2\right]\nonumber\\
&&\simeq\bar{\psi}i\slashed{D}{\psi}-\sigma(r)(\bar{\psi}{\psi})-\frac{1}{2G}\sigma^2+\psi^\dagger\vec{\omega}\cdot\vec{J}\psi
\end{eqnarray}
where quantities with tilde are corresponding to those in the curved metrics while ordinary ones are those in the flat spacetime. In the cylindrical symmetric system the mean field chiral condensate $\sigma(r)=-G\langle\bar\psi \psi\rangle$ shall generally depend on the radial coordinate $r$ explicitly in a rotating system.

\begin{figure}[!hbt]
\begin{center}
\tikzstyle{startstop} = [rectangle, rounded corners, minimum width=1cm, minimum height=1cm,text centered, draw=black, fill=red!30]
\tikzstyle{io} = [trapezium, trapezium left angle=70, trapezium right angle=110, minimum width=1cm, minimum height=1cm, text centered, draw=black, fill=red!30]
\tikzstyle{process} = [rectangle, minimum width=1cm, minimum height=1cm, text centered, draw=black, fill=orange!30]
\tikzstyle{decision} = [diamond, aspect=1.2, minimum width=1cm, minimum height=1cm, text centered, draw=black, fill=green!30]
\tikzstyle{arrow} = [thick,->,>=stealth]

\begin{tikzpicture}[node distance=2cm]

\node (in1) [io] {Step 0: Input $\sigma_0=\sigma_{spec}$};
\node (pro1) [process, below of=in1] {Step i: Solve $H[\sigma_i(r)]\psi_n^i=E_n \psi_n^i$};
\node (dec1) [decision, below of=pro1, yshift=-1cm, align=center] {$\sigma^{i+1}=\langle\bar\psi^i \psi^i\rangle$\\\\$||\sigma^{i+1}-\sigma^i||<\epsilon$ };
\node (pro2b) [process, right of=dec1, xshift=1.5cm, align=center] {Update\\ $H[\sigma^{i+1}]$};
\node (out1) [io, below of=dec1, yshift=-1cm] {Output $\sigma_{final}(r)$};

\draw [arrow](in1) -- (pro1);
\draw [arrow](pro1) -- (dec1);
\draw [arrow](dec1) -- (out1);
\draw [arrow](dec1) -- (pro2b);
\draw [arrow](dec1) -- node[anchor=east] {Yes} (out1);
\draw [arrow](dec1) -- node[anchor=south] {No} (pro2b);
\draw [arrow] (pro2b) |-(pro1);
\node at (3.0, -3.5) {i++};
\end{tikzpicture}
\end{center}
\caption{ The flowchart of the self-consistent framework.}
\label{fig_0}
\end{figure}
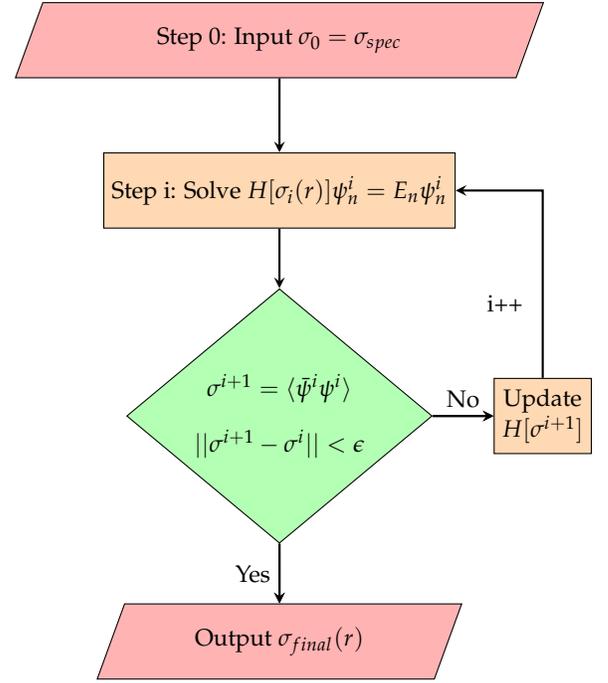

Noting that for a position-dependent angular velocity the linear velocity should be solved by the definition $\vec\omega=\frac{1}{2}\nabla\times \vec{v}$ whose solution is $v(r)=2 r^{-1}\int_0^r d\rho \rho \omega(\rho)$ along the angular direction. Obviously the Hamiltonian is a functional of the undetermined chiral condensate $\sigma(r)$. In order to solve it self-consistently the BdG-like algorithm is adopted as Fig. \ref{fig_0} whose general idea is almost the same as that in the introduction. In the relativistic version the Sch\"{o}rdinger equation shall be replaced by Dirac equation. And a larger basis shall be chosen to expand a spinor with 4 components. In our computation the Manhattan norm $||f(r)||=\frac{2}{R^2}\int_0^Rdr r |f(r)|$ is used. If convergence constraint is chosen as $\epsilon=10^{-2}$ in this report, the solution could usually be found in dozens of steps.

During this process the set of Bessel functions of the 1st kind, which serves as a suitable complete basis in a cylindrical symmetric system, is adopted to expand those four components of each eigenstate of spinor $\psi_n=(u^{l,\uparrow}_{n}, u^{l,\downarrow}_{n}, v^{l,\uparrow}_{n}, v^{l,\downarrow}_{n})^T$ as
\begin{gather}
u^{l,\uparrow}_{n}=\frac{c^{\uparrow}_{n,j}}{\sqrt{2\pi}}\varphi_{j,l}(\rho)e^{i(l+1)\theta},\
u^{l,\downarrow}_{n}=\frac{c^{\downarrow}_{n,j}}{\sqrt{2\pi}}\phi_{j,l}(\rho)e^{i l\theta}\nonumber\\
v^{l,\uparrow}_{n}=\frac{d^{\uparrow}_{n,j}}{\sqrt{2\pi}}\chi_{j,l}(\rho)e^{i(l+1)\theta},\
v^{l,\downarrow}_{n}=\frac{d^{\downarrow}_{n,j}}{\sqrt{2\pi}}\xi_{j,l}(\rho)e^{il\theta}
\end{gather}
where repeated indices are implicitly summed over and $ \varphi_{j,l-1}(\rho)=\phi_{j,l}(\rho)=\chi_{j,l-1}(\rho)=\xi_{j,l}(\rho)=\frac{\sqrt{2}J_l(\alpha_{j,l}\frac{\rho}{R})}{RJ_{l+1}(\alpha_{j,l})} $, $ \alpha_{j,l} $ is the $ j$-th zero point of $ J_l(x) $ and we use the up(down) arrow to tag the spin label and $n$ for energy levels. $R$ is the radius of the system.

According to \cite{Rosenstein:1990nm} the coupling renormalization scheme $G(\Lambda)=\pi/(\Lambda-M_0)$ is adopted, where $\Lambda$ is the energy threshold of the energy level summation\footnote{It has been checked the numerical results do not depend on the $\Lambda$ because of the renormalizability}.
\begin{eqnarray}
\sigma(r)=-G(\Lambda)\sum_{|E_n|<\Lambda}[\bar\psi_n(r)\psi_n(r)][\frac{1}{2}+f_{FD}(E_n)]
\end{eqnarray}
where $f_{FD}$ is the Fermi-Dirac distribution. And $M_0$ is the value of chiral condensate in the vacuum, i.e. the infinite and static system at zero temperature. In the following all of the quantities are in unit of $M_0$. In a static system, numerically it is easy to confirm that if the system radius is comparable with $1 M_0^{-1}$, which could be treated roughly as a typical correlation length of the system, the condensate would be larger than a unit at zero temperature. This is known as the usual finite-size effect. And it will decrease towards a unit as radius increasing at $\omega=0$. It is found that when $R>10M_0^{-1}$ the condensate will be almost a unit everywhere except the boundary. Hence in order to study inhomogeneous effects in an infinite rotating system we choose a relatively large radius $R=30M_0^{-1}$ to mimic the infinite case in the following. Even in such a large system there is still remarkable oscillation at the boundary because of the boundary condition $\psi(r=R)=0$. However it could be ignored safely here. Because we have carefully checked that both the inner condensate profile and the chiral condensate behavior with temperature have already restored to those in the vacuum and will not be influenced by the boundary condition in the system with $R>30M^{-1}_0$. And as complement we also check in this framework all the LDA results could be regenerated. The detailed boundary and system size dependence results as well as the uniformly rotating system studies will be reported in our future papers.

{\it Local suppression effect---}
In the LDA, i.e. assuming $\partial^n\sigma/\partial r^n\simeq 0$ for $n>0$, the condensate $\sigma(r)$ is treated as a constant in each local {\it cell}. It is found that rotation behaves as a different kind of "effective" chemical potential $\mu_\omega=(l+\frac{1}{2})\omega$ in the phase diagram\cite{Jiang:2016wvv}. It is easy to understand by noting the $J_z$ is a quantum mechanical operator which means $[H, J_z]=0$ if $\partial\omega/\partial r=0$\cite{Jiang:2016wvv}. In this approximation the impact of boundary condition has also been studied under the constraint of the light speed limit. It is found that at zero temperature the rotation speed limit lead to a slightly surprised fact that the uniform rotation will have no impact on the condensate in a finite-size system with the sharp boundary condition $\psi(R)=0$\cite{Ebihara:2016fwa}. It is a natural conclusion by noticing eigen energies $E_{l j}=\alpha_{l j}/R$ and $\omega<1/R$ lead to $E_{l 1}>(l+1/2)\omega$ holding for any $l$. However the LDA is just a palliative to solve the Dirac equation analytically, because it is only valid in those cases in which the rotation speed is too slow to bend the constant condensate profile in the vacuum remarkably. In this point of view this approximation practically is too limit to explore rotation effects on the condensate structure self-consistently, especially in inhomogeneous cases.  Nevertheless these studies indicate that we could roughly estimate the influence of rotation by considering it as a effective chemical potential. This estimation will be used in the following rotation speed choices.

\begin{figure}[!hbt]
\begin{center}
\includegraphics[scale=0.45]{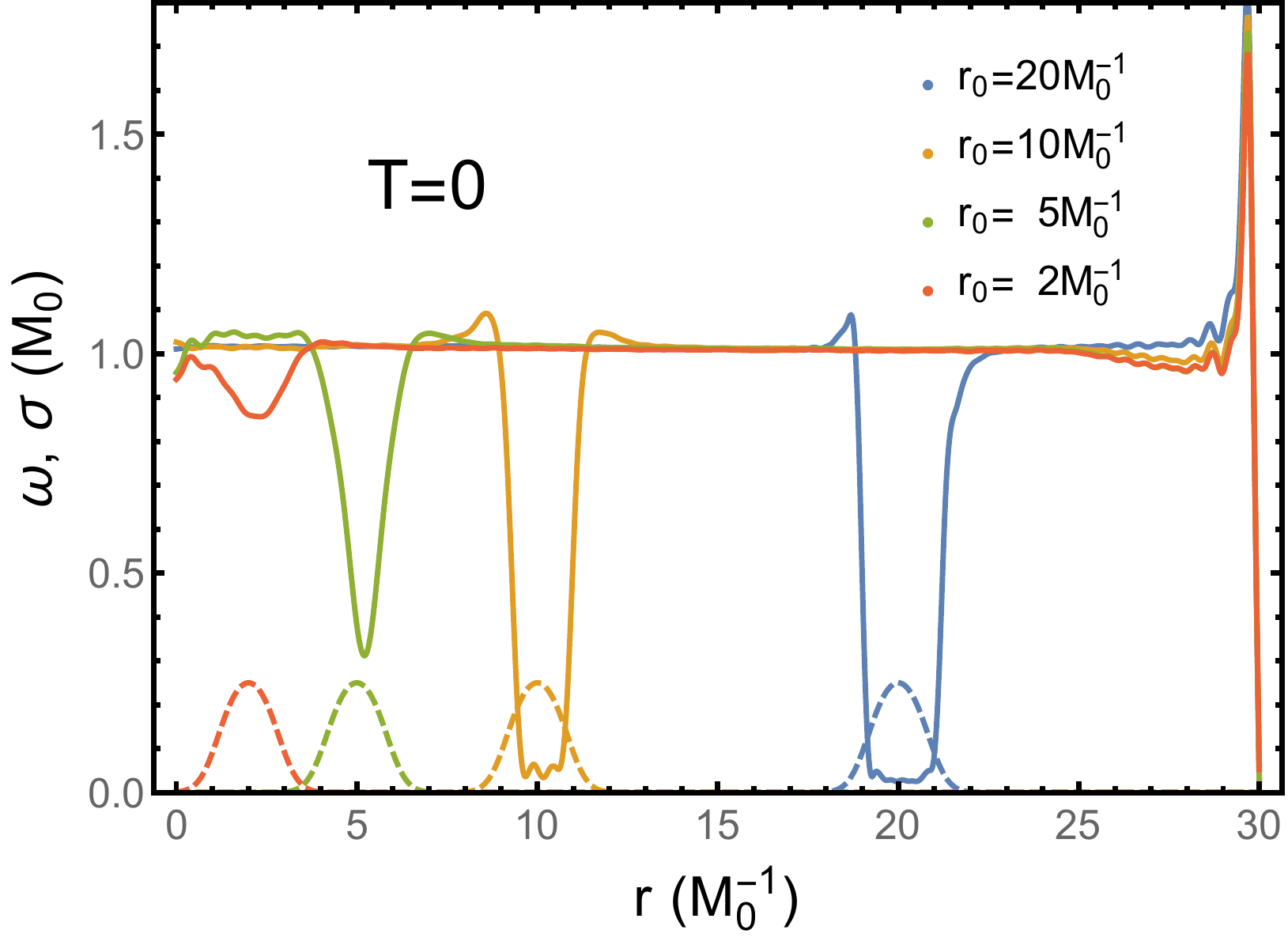}
\caption{ The chiral condensate $\sigma(r)$ profiles(solid lines) at zero temperature under different rotations $\omega=0.5[exp(1.5(r-r_0)^2)+1]^{-1}$ (dash lines).}
\label{fig_1}
\end{center}
\end{figure}

Although code-tuning a little tedious, in the relativistic BdG framework it is straightforward to compute the condensate profile under rotation self-consistently. In order to go beyond the LDA case we firstly choose the rotation profile according to two constraints as follows. First an inhomogeneous and decaying rotation is necessary. Because for an uniform case the $\omega\leq R^{-1}=1/30 M_0$ is much less than the vacuum mass $M_0$ of fermions. According to the effective chemical potential estimation it is too small to induce visible effects. Secondly, the linear velocity expression suggests the large rotation speed range should not be too wide to exceed the speed of light. Hence a gaussian-like or $\delta$-function distribution whose peak locating at some finite radial coordinate is the simplest choice. In our work we adopt the gaussian-like one for numerical computation and fix the peak value as $0.25M_0$ which is large enough to result in significant modifications on chiral condensate. The other parameters in the rotation profile are adjusted to maintain the maximum linear velocity consistent with the metrics expansion approximation. The chiral condensate profiles at zero temperature under different rotations are obtained in Fig.\ref{fig_1}.

In Fig.\ref{fig_1} it is clear that even at zero temperature condensates are significantly modified by rotations. They all stay as the vacuum value, i.e. a unit, in the non-rotating ranges, while suppressed in the finite rotating sections. As works in LDA suggested the rotation shall suppress the chiral condensate, because for this scalar pairing state, one of the relative orbital angular momentum and their individual spins shall be opposite to the rotation direction. Hence the polarization effect of rotation $\vec{J}\cdot\vec{\omega}$ tends to destroy this kind of pairing. Besides this well-known qualitative behavior, there are two interesting properties in this result. Firstly the suppression effect emerges locally and locates in the section where the rotation is maximum. This is actually not so surprising in consideration of the typical correlation length is around $1M_0^{-1}$ which almost equals to the wide of rotation speed peak. This will weaken the impact of the rotation peak on the distant part, thus lead to the local suppression effect which is the direct consequence of the rotational polarization effect. It is worthwhile to mention that this local suppression effect could also be understood by the {\it local} distortion of the corresponding energy levels. This will also be reported in our future paper. In the meantime an interesting question may arise: is it possible to set a rotation profile whose peak could be much wider than the typical correlation length? The answer is negative. We could estimate the width by choosing the rotation as $\omega(r)=\omega_0$ in the section $r\in[r_0, r_0+\delta r]$. Obviously the $\delta r$ shall not be too large because of the speed limit. By assuming $\delta r << r_0$ the constraint on the maximum linear velocity is solved as $v_{max}=2\omega_0\delta r < 1$. As the $\omega_0$ is supposed to be larger than around $0.2M_0$ to induce significant suppression effect, thus the $\delta r$ could not exceed around $2M^{-1}_0$. This indicates it would be always valid to estimate the rotational suppression effect locally.

Secondly the radial coordinate will enhance the rotational suppression effect. It appears that it is the combination $\omega r$ rather than the angular velocity itself determines the rotational contribution of suppression. Actually it has already been discovered in our previous work \cite{Jiang:2016wvv}. It is easy to understand by noting that at larger $r$ the orbital angular momentum becomes larger as well at the same $\omega$ and transverse momentum scale. As a result the condensate at such position would experience more suppression effect. Although in this nonuniform rotating system the transverse momentum $k_t$ is no longer a quantum mechanical operator, it is still intuitive to understand the $\omega r$ in such a way by noting the distortion of energy levels and condensate suppression are both finite and local.

{\it Centrifugal effect.---}
It has been shown that the fast-enough rotation will suppress the chiral condensate locally because of the rotational polarization. What about slower ones? Is it simply negligible according to the effective chemical potential estimation? The answer is negative. When the magnitude of the rotation is small enough, its first derivative would play a more important role. In the following we adopted a Woods-Saxon-shaped rotation profile whose saturated value is $0.06M_0$. The plateau part covers $10M_0^{-1}$ from the center and decreases exponentially towards zero at large distance.

\begin{figure}[!hbt]
\begin{center}
\includegraphics[scale=0.45]{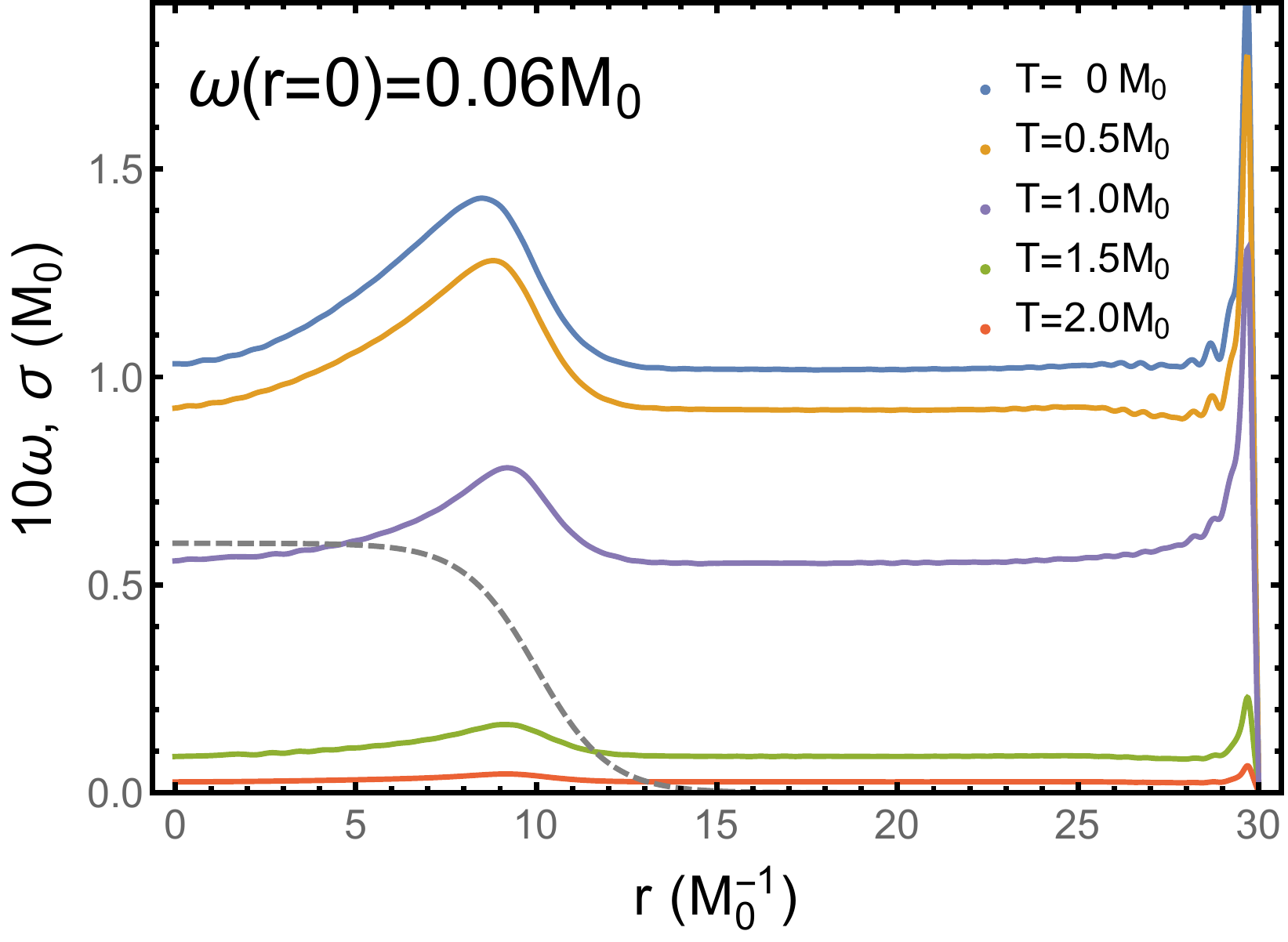}
\caption{ The chiral condensate profile $\sigma(r)$ (solid lines) at different temperatures under rotation $\omega=0.06[exp(r-10)+1]^{-1}$(dashed line).}
\label{fig_2}
\end{center}
\end{figure}

Fig.\ref{fig_2} shows the corrresponding chiral condensates at different temperatures. Focusing on the zero-temperature case there are two properties for the slow-rotating case. Firstly the condensate starts from a unit and decreases back to it along with rotation at large distance. This means the rotation is too slow to induce the suppression effect. The second property is a behavior beyond the local rotational suppression effect. Instead of a valley there is a bump in the radial section with finite rotation. It increases slowly in the range of the rotation plateau and falls back to a unit rapidly along with the rotation. Intuitively we will refer to this behavior as the centrifugal effect analogy to the picture of most classical systems, such as the surface of rotating water. As the name indicates, this phenomenon shall be a generic one which has also been observed in cold atom systems\cite{Bausmerth:2008}. This phenomenon is driven by the "force" along the radial direction which could be explicit derived in the Heisenberg equation\cite{Anandan:2003}. In such a slowly rotating system, technically the condensate profile is supposed to be mainly induced by the $\partial\omega/\partial r$. Furthermore by taking the degeneracy of $\pm\omega$ contributions into account, the conclusion could be refined as the centrifugal effect is actually driven by $|\partial\omega/\partial r|$. This is exactly the reason why in the Fig.\ref{fig_1} there is a small bump at both ends of the square well of the condensate where values of $|\partial\omega/\partial r|$ are maximum. As the temperature increasing, the chiral condensate will be melted as expected. At around $T\simeq 2M_0$ the chiral symmetry has almost been restored. This critical temperature is consistent with the result in \cite{Appelquist:1986fd} with the same model as our work.

{\it Summary and Discussions.---}
In this work we developed a self-consistent and relativistic framework to study the inhomogeneous condensate of fermion pairing under rotation. This work is necessary because rotating systems shall involve inhomogenous phenomena inevitably. Focusing on the chiral condensate the local suppression and centrifugal effects have been discovered within this framework. As a generic phenomenon the rotation induced inhomogeneousness would play important roles in various physics systems. In HIC these results suggest that the nonuniform chiral condensate may have remarkable impact on the QGP-fireball evolution, in which there are both large local vorticity and drastic spatial fluctuations. According to one of HIC simulations the largest local vorticity could reach 0.1$\sim$0.2GeV at the early stage of collision, which is comparable to the component quark mass 0.3GeV\cite{Jiang:2016woz}. This indicates that the phase transition process in the early stage of the collision might be much more than the scenario of temperature driven chiral restoration. For example in the cooler part but carrying larger vorticity, rotational suppression effect could speed up the chiral restoration, which would have been difficult to take place with its own temperature. While in the area with medium temperature the centrifugal effect would slow down the chiral restoration because of the large derivative there. Hence this naive analysis indicates chiral condensate distribution, which induced by temperature fluctuation at corresponding positions, would be further smoothed by the inhomogeneous rotational effects. This modification on the phase transition process may improve current pictures of both thermalization and freeze-out. Aiming at these phenomenological topics we are focusing on making use of this framework to study the phase transition process under a more realistic vorticity distribution as well as combining this static framework with a transport model. For the nuclear matter in neutron stars besides the rotational suppression effect, the centrifugal effect may play a more interesting role in the pulsar glitch behavior. If it could be understood by the coupling of the pulsar's faster-spinning superfluid core to the crust\cite{Link:1992mdl,Melatos:2007px}, centrifugal effect will introduce another non-negligible contribution to the structure studies of neutron stars. Besides the scalar pairing state more novel states could exist in the rotating system. The simplest one is called the vortex state which has been discovered in  non-relativistic cold atom systems, e.g. in\cite{Haljan:2001zz,Eto:2013hoa}. As a generic framework of studying fermion pairings it is also straightforward to search such states in the relativistic version. We will report more details of this work in another paper.

\vspace{0.2in}

 {\bf Acknowledgments.}
The authors thank K. Fukushima and X.G. Huang for helpful discussions and comments.  The work of this research is supported by the National Natural Science Foundation of China, Grant No. 11575093(LX and PZ), 11890712(LX, PZ and LH), 11775123(LH), 11875002(YJ) and the National Key R\&D Program of China(Grant No. 2018YFA0306503)(LH). LX and YJ are also supported by the China Scholarship Council(CSC) for visiting at the University of Tokyo and the Zhuobai Program of Beihang University respectively.

\end{document}